\pdfoutput=1
\documentclass[12pt]{article}

\usepackage{graphicx}
\usepackage{float,amsmath}
\usepackage{natbib,enumitem,multirow}

\title{Mean square displacement and instantaneous diffusion coefficient of charged particles in stochastic motion}
\author{Gabriela Raluca Mocanu\footnote{Astronomical Observatory Cluj, Romanian Academy, Cluj-Napoca Branch, 19 Ciresilor Street, 400487, Cluj-Napoca, Romania,
Email: gabriela.mocanu@academia-cj.ro}}
\date{}
\begin{document}
\maketitle
\noindent \textbf{keywords}\\  radiation: dynamics-- methods: numerical

\begin{abstract}
The mean square displacement and instantaneous diffusion coefficient for different configurations of charged particles in stochastic motion are calculated by numerically solving the associated equations of motion. The method is suitable for obtaining accurate descriptions of diffusion in both intermediate and long time regimes. It is also appropriate for studying a variety of astrophysical configurations since it may incorporate microscopic physics that analytical methods cannot cope with. The results show that, in the intermediary time regime, the diffusion coefficient has an irregular behavior, which can be described in terms of the complex interplay appearing between the physical parameters describing the configuration. The main conclusion is that such an approach may serve at differential diagnosis of different astrophysical configurations.
\end{abstract}

\section{Introduction}

In an astrophysical context, there is overwhelming observational evidence that the plasma emitting the recorded radiation is influenced by some stochastic component in its medium (for discussion of observational data and methods that lead to this conclusion for, e.g., accretion disks around supermassive black holes see~\cite{aza05,car11,leu11,har14}). As this plasma moves, it emits radiation, which is our only source of information for diagnosis of the emitting system; we find it appropriate to describe this complicated framework by instantaneous quantities, such as the instantaneous diffusion coefficient. This approach takes into account all the regimes of the system evolution; the transitory regime is worth studying as it may be used to model and diagnose explosive astrophysical events.

It is thus the purpose of this paper to use the advancements in numerical computation to study the instantaneous diffusion coefficient by solving the stochastic differential equation of motion, or sometimes called the A-Langevin equations, associated to the motion of the particles in the plasma.

The trajectory $x(t)$ of a particle undergoing a random motion is a solution of the stochastic differential equation (SDE)
\begin{equation*}
\frac{d^2x(t)}{dt^2} = a_{det}+a_{stoch},
\end{equation*}
where $a_{det}$ is the deterministic contribution to the equation of motion and $a_{stoch}$ is the stochastic contribution to the equation of motion, such as random collisions causing momentum variations. The mean square displacement (msd) $\langle x^2(t)\rangle$ exhibited by an ensemble of such particles can be numerically calculated by solving the associated equation of motion. The retrieval of the msd is necessary because the (macroscopic) instantaneous diffusion coefficient is defined as~\cite[Eq. 6]{bal94}
\begin{equation}
D(t) = \frac{1}{2}\partial _t \langle x^2(t)\rangle \label{eq:coeffDef},
\end{equation}
with the equilibrium diffusion coefficient defined as
\begin{equation*}
D = \lim _{t \to \infty} D(t).
\end{equation*}

In the numerical treatment the motion will be split into a mean and a fluctuating part $x(t) = \bar{x}(t) + \delta x (t)$, with $\delta x (0) = \delta v (0) = 0$. \emph{From hereon, the diffusion coefficient will thus study the diffusion with respect to a mean trajectory $\bar{x}(t)$}, where this mean trajectory is the solution to the deterministic equation of motion.

The diffusion coefficient will obviously inherit the properties of both types of accelerations appearing in the equation of motion; these properties are in turn signatures of the underlying physics of the particle-medium interaction. Since analytical calculation of these properties, by using direct integration of the SDEs, are not generally possible, a numerical integrator allows to investigate the importance of complex microscopic interactions. We study four physical configurations of charged particles undergoing random collisions:
\begin{enumerate}[label=(\Alph*)]
\item{}in a constant electric field with constant friction;
\item{}in an external harmonic potential with constant friction;
\item{}in an external harmonic potential while subjected to friction with memory;
\item{}in a constant magnetic field with constant friction.
\end{enumerate}

These cases were previously investigated in~\cite{har16}, where results regarding the radiation of the charged particle were obtained.

To numerically investigate the trajectories of the charged particle in each of these cases, different suitable numerical methods were employed for each case: Euler methods for cases (A) and (B), a second order Runge-Kutta method developed in~\cite{hers} for case (C) and a combination between an Euler method and the method developed in~\cite{b4} for case (D). The technical approach may be largely described by the successive steps: 1) establish the appropriate equation of motion, 2) make it dimensionless and 3) numerically solve it (as detailed in Section~\ref{sect:method}).

The main results of the investigation may be split into two main areas: the intermediate behavior of $D(t)$ and its long time limit $D$. For both cases, the results depend on the parameters of the systems, most notably on the 1) ratio between the stochastic acceleration and the acceleration of the particle between collisions, and 2) on the ratio between the stochastic acceleration and the effective frequency of the external harmonic potential. The results are detailed in Section~\ref{sect:res}.

The numerical value of the long time limit of $D(t)$ is a constant. But the intermediate time regime of $D(t)$ is more interesting as it shows non-trivial behavior. As a general conclusion, we state that there are time intervals (with onset and length characteristic to each parameter configuration) in which diffusion is extremely strong as compared to the long time limit. This raises questions of diagnosis of system parameters by means of observational data and even more, issues of confinement. A more detailed discussion of the results is given in Section~\ref{sect:disc}.

\section{Possible settings for random motion}\label{sect:method}

\subsection{Stochastic motion in the presence of an electric field}

The simplest possible stochastic motion of a charged particle with $Z=1$ and
mass $m$ is the one-dimensional Brownian motion in the presence of an
external electric field, $\vec{E}\neq 0$. In the following we restrict or analysis to the case of the constant electric field, $\vec{E}={\rm constant}$. The random motion of the particle is described by the Langevin equation
\begin{equation}  \label{Lang1}
\frac{d^2x}{dt^2} = eE- \nu \frac{dx}{dt} + \xi _A(t),
\end{equation}
where $\xi _A $ is a random acceleration with properties defined through the ensemble averages~\citep{bal97},
\begin{equation}
\left \langle \xi _A (t)\right \rangle =0, \quad \left \langle \xi _A
\left(t_1\right)\xi _A \left(t_2\right)\right \rangle = A\delta
\left(t_1-t_2\right).
\end{equation}

The physical quantities are split into their mean and fluctuating parts, and thus we obtain the SDE for $\delta x$ as
\begin{equation}  \label{Lang1}
\frac{d^2\delta x}{dt^2} = - \nu \frac{d \delta x}{dt} + \xi _A(t).
\end{equation}

The differential equation is brought to a dimensionless form by the following transformations

\begin{itemize}
\item dimensionless time: $\theta = \nu t$; $\nu= 1/\tau$, where $\nu$ is
the collision frequency in Brownian Motion;

\item dimensionless fluctuation in displacement: $q = \delta x \left ( A \tau ^3 \right ) ^{-1/2}$.
\end{itemize}

The dimensionless equation describing the motion of a Brownian particle in a constant electric field is
\begin{equation}
\frac{d^2q (\theta)}{d\theta ^2} = - \frac{dq}{d\theta} (\theta) + \bar{\xi}_A (\theta),\label{eq:velA}
\end{equation}
where
\begin{equation}
\left \langle \bar{\xi}_A(\theta)\right \rangle =0, \quad \left \langle \bar{\xi}
_A \left(\theta_1\right)\bar{\xi} _A \left(\theta_2\right)\right \rangle=%
\bar{A}\delta \left(\theta_1-\theta_2\right).\label{eq:noiseA}
\end{equation}

Note that although the constant electric field is important to the overall energy emission~\citep{har16}, it does not change the characteristics of the fluctuating component of the trajectory and thus it will not affect the diffusion.

The free parameter set in this case is given by $\{ \bar{A}\}$.

\subsection{Stochastic motion in a harmonic external potential}

The Langevin equation for the one dimensional motion of a charged particle with mass $m$
and charge $e$ in a harmonic potential with natural frequency $\omega _{0}$
is given by
\begin{equation}
\frac{d^{2}x}{dt^{2}}+\nu \frac{dx}{dt}+\omega _{0}^{2}x=\xi _{B},
\label{l1}
\end{equation}%
where the stochastic acceleration $\xi _{B}$ has the properties
\begin{equation}
\left\langle \xi _{B}(t)\right\rangle =0, \quad \left\langle \xi _{B}\left(
t_{1}\right) \xi _{B}\left( t_{2}\right) \right\rangle =B\delta
\left( t_{1}-t_{2}\right) .
\end{equation}

Thus the fluctuating part of the trajectory will obey the SDE
\begin{equation}
\frac{d^{2}\delta x}{dt^{2}}+\nu \frac{d\delta x}{dt}+\omega _{0}^{2}\delta x=\xi _{B}.
\label{l1}
\end{equation}

The same dimensionless variables are again used, together with
\begin{itemize}

\item {} dimensionless frequency: $W=\omega _0 \tau$.

\end{itemize}

The dimensionless form of the Langevin differential equation~(\ref{l1}) becomes
\begin{equation}\label{harmosc}
\frac{d^2q (\theta)}{d\theta ^2} + \frac{dq(\theta)}{d\theta} + W^2 q
(\theta)= \bar{\xi} _B (\theta),
\end{equation}
where
\begin{equation}
\left \langle \bar{\xi}_B(\theta)\right \rangle =0, \quad \left \langle \bar{\xi}
_B \left(\theta_1\right)\bar{\xi} _B \left(\theta_2\right)\right \rangle=%
\bar{B}\delta \left(\theta_1-\theta_2\right).\label{eq:noiseB}
\end{equation}

In this case, the characteristics of the outside medium (through $W^2$) do influence the msd and thus the diffusion coefficient.

The free parameter set in this case is given by $\{W^2, \bar{B}\}$.

\subsection{Stochastic motion described by the generalized Langevin equation}

In the presence of a non-trivially correlated noise and of a frictional force showing retarded effects, the motion of a charged particle in a harmonic potential is described by the generalized Langevin equation, which in the one-dimensional case can be written as (see, e.g., \cite{har16})
\begin{equation}\label{l2}
\frac{d^2 x}{dt^2}+\int_0^t{\gamma (t-t^{\prime }) \frac{dx\left(t'\right)}{dt^{\prime }}}%
dt^{\prime  }+\omega _0^2x=\xi _C (t),
\end{equation}
where
\begin{equation}
\gamma (t) = \frac{\alpha}{\tau _d} \exp \left \{ -t/\tau\right \}, \quad \langle \xi _C (t) \xi _C (t^{\prime }) \rangle = \frac{1}{\beta} \gamma
(t-t^{\prime }).
\end{equation}

The SDE for the fluctuating part of the trajectory is
\begin{equation}\label{l2}
\frac{d^2 \delta x}{dt^2}+\int_0^t{\gamma (t-t^{\prime }) \frac{d \delta x\left(t'\right)}{dt^{\prime }}}%
dt^{\prime  }+\omega _0^2 \delta x=\xi _C (t).
\end{equation}

The previously defined dimensionless quantities are used, together with
\begin{itemize}
\item {} dimensionless friction amplitude: $\bar{\alpha} = \alpha \tau$;

\item {} dimensionless friction kernel: $\bar{\gamma} (\theta) = \bar{\alpha}
e^{-\theta}$;

\item {}dimensionless correlation amplitude for the stochastic force $\bar{C}=
\alpha \nu/\beta (\nu v_T)^{-2}$.
\end{itemize}

The dimensionless equation becomes
\begin{equation}
\frac{d^2q}{d\theta ^2} + \int _0 ^\theta \bar{\gamma} (\theta - \theta
^{\prime }) \frac{dq}{d\theta ^{\prime }}d\theta ^{\prime 2 } + W^2 q = \bar{\xi}_C
(\theta),\label{eq:velC}
\end{equation}
where
\begin{equation}
\langle \bar{\xi} _C (\theta) \bar{\xi} _C (\theta ^{\prime }) \rangle = \bar{C}
e^{\theta ^{\prime }- \theta}.\label{eq:noiseC}
\end{equation}

For the technical details on how to solve this type of equation, see~\cite{har16}.

The free parameter set in this case is given by $\{\bar{\alpha}, \bar{C}, W^2\}$.

\subsection{Stochastic motion in a constant magnetic field}

The equation of motion of a charged particle in a magnetic field in the presence of a stochastic acceleration  $\vec{\xi} ^D(t)$ and of interparticle collisions, generating an acceleration proportional to the particle velocity, is given by the Langevin type equation~\citep{b4}
\begin{equation}\label{l4}
\frac{d\vec{v}}{dt} = \frac{Ze}{mc}\left [ \vec{v}(t)\times \vec{B}\right ]
- \nu \vec{v}(t) + \vec{\xi} ^D(t)
\end{equation}
where
\begin{equation}
\left \langle \xi ^D_i \left(t\right)\xi ^D _j \left(t^{\prime
}\right)\right \rangle = D \delta _{ij} \left(t-t^{\prime
}\right), \quad i,j=x,y
\end{equation}
and we consider a constant magnetic field oriented along the $z$ direction,
\begin{equation}
\vec{B} = B\hat{z},B={\rm constant}.
\end{equation}

In addition to the dimensionless quantities defined so far, we define a
\begin{itemize}
\item dimensionless magnetic frequency: $\bar{\Omega} = \Omega \tau = B_0%
\frac{Ze}{mc}\tau$.
\end{itemize}

Equation~(\ref{l4}) is split into components and afterwards
made dimensionless as
\begin{equation}
\quad \frac{d^2 X}{d\theta ^2}=\bar{\Omega}\frac{dY}{d\theta} - \frac{dX}{d\theta} + \bar{\xi}_x (\theta),\label{eq:DX}
\end{equation}
\begin{equation}
\quad\frac{d^2Y}{d\theta ^2}=-\bar{\Omega}\frac{dX}{d\theta} - \frac{dY}{d\theta} + \bar{\xi}_y (\theta),\label{eq:DY}
\end{equation}
\begin{equation}
\quad\frac{d^2 Z}{d\theta ^2}=- \frac{dZ}{d\theta} + \bar{\xi}_z (\theta),
\end{equation}
where
\begin{equation}
\left \langle \bar{\xi} _i \left(\theta \right)\bar{\xi} _j \left(\theta
^{\prime }\right)\right \rangle =\bar{D}\delta _{ij} \left(\theta
- \theta ^{\prime }\right). \label{eq:noiseD}
\end{equation}

The free parameter set in this case is given by $\{\bar{\Omega}, \bar{D}\}$.

\section{Results}\label{sect:res}

The stochastic differential equations \eqref{eq:velA}-\eqref{eq:noiseA}, \eqref{harmosc}-\eqref{eq:noiseB}, \eqref{eq:velC}-\eqref{eq:noiseC}, \eqref{eq:DX}-\eqref{eq:noiseD} were solved for $q(\theta)$, subsequently used to produce the time evolution of the mean square displacement and the instantaneous diffusion coefficient. Simulations were run for $10^5$ realizations, $10^3$ timesteps each, within an extended parameter set (Table~\ref{tab:params}).

\begin{table}
	\centering
	\caption{Parameter space tested by the simulations. In the first column the case is specified, in the second the parameter space tested in the simulations and the third column gives the total number of different settings covered by the simulations.}
	\label{tab:params}
	\begin{tabular}{p{0.08\linewidth}| p{0.5\linewidth}| p{0.08\linewidth}}
		\hline
		Case & Parameter space & No.\\
		\hline
		(A) & $\bar{A}\in \{0.01, 1, 100\}$ & $3$ \\ \hline
        (B) & $\bar{B} \in \{0.01, 1, 100\}$, $W^2 \in \{0.01, 0.05,0.1,0.5,1,5,10,20 \}$ & $3\times 8$ \\ \hline
        (C) & $\bar{C}\in \{0.01, 1, 100\}$, $W^2 \in \{0.01, 0.05,0.1,0.5,1,5,10,20 \}$, $\bar{\alpha} \in \{0.01, 0.05,0.1,0.5,1,5,10,20 \}$ & $3\times 8 \times 8$ \\ \hline
        (D) & $\bar{D}\in \{0.01, 1, 100\}$, $\bar{\Omega} \in \{0.01, 0.05,0.1,0.5,1,5,10,20 \}$ & $3 \times 8$\\ \hline
	\end{tabular}
\end{table}

\begin{figure}[H]
\begin{center}
	\includegraphics[width=0.8\textwidth]{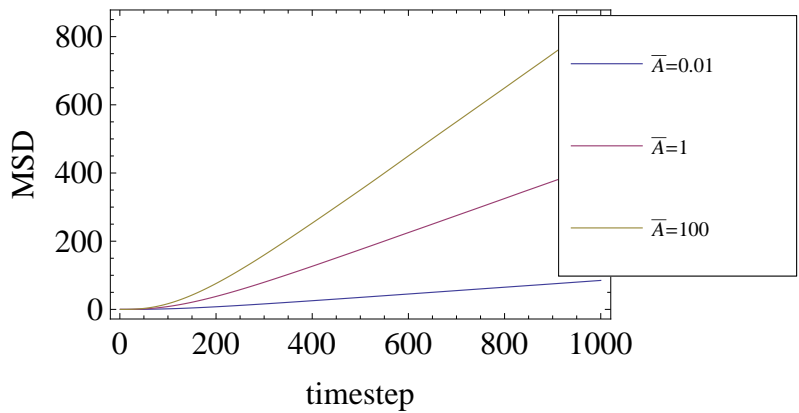}
    \caption{Mean square displacement of the charged particle undergoing Brownian Motion in a constant electric field (A), for various values of the noise amplitude $\bar{A}$; for presentation purposes the curves were multiplied by $10^7$ for the $\bar{A}=0.01$ curve and by a factor of $5000$ for the $\bar{A}=1$ curve.}
    \label{fig:AMSD}
\end{center}
\end{figure}

\begin{figure}[H]
\begin{center}
	\includegraphics[width=0.8\textwidth]{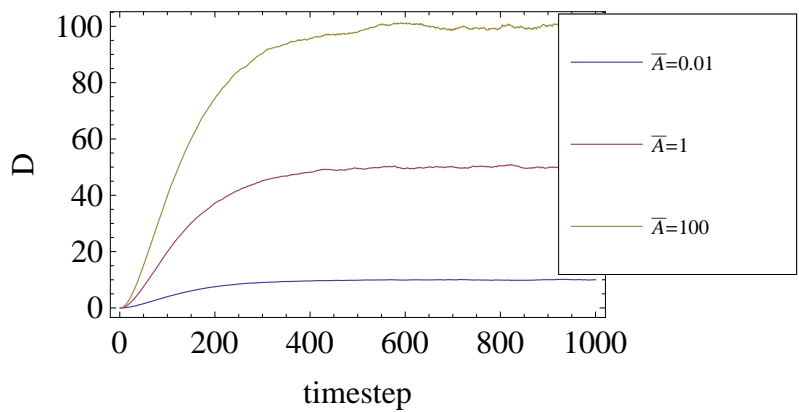}
    \caption{Diffusion coefficient for case (A) (msd shown in Fig.~\ref{fig:AMSD}); for presentation purposes the curves were multiplied by $10^7$ for the $\bar{A}=0.01$ curve and by a factor of $5000$ for the $\bar{A}=1$ curve.}
    \label{fig:ADiff}
		\end{center}
\end{figure}

\begin{figure}[H]
\begin{center}
	\includegraphics[width=0.8\textwidth]{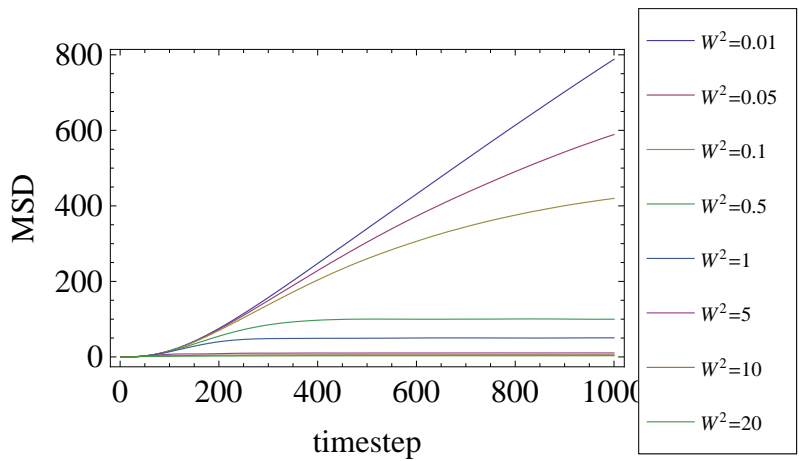}
    \caption{Mean square displacement of the charged particle undergoing stochastic motion in an external harmonic potential (B), for $\bar{B}=100$.}
    \label{fig:BMSD}
				\end{center}
\end{figure}

\begin{figure}[H]
\begin{center}
	\includegraphics[width=0.8\textwidth]{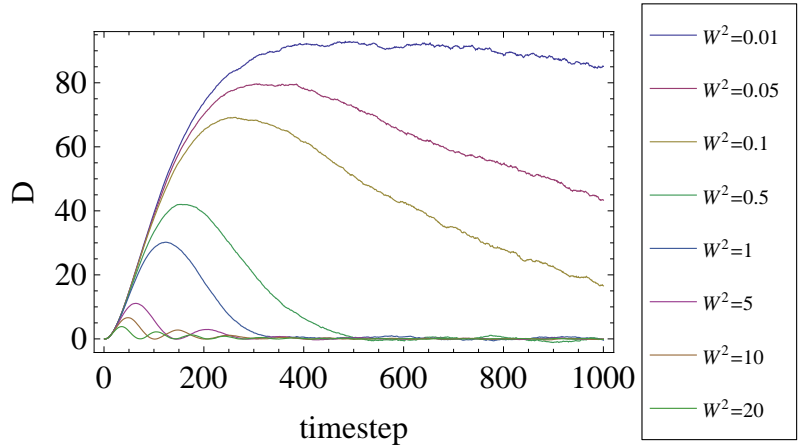}
    \caption{Diffusion coefficient for case (B) (msd shown in Fig.~\ref{fig:BMSD}), for $\bar{B}=100$.}
    \label{fig:BDiff}
				\end{center}
\end{figure}

\begin{figure}[H]
\begin{center}
	\includegraphics[width=0.8\textwidth]{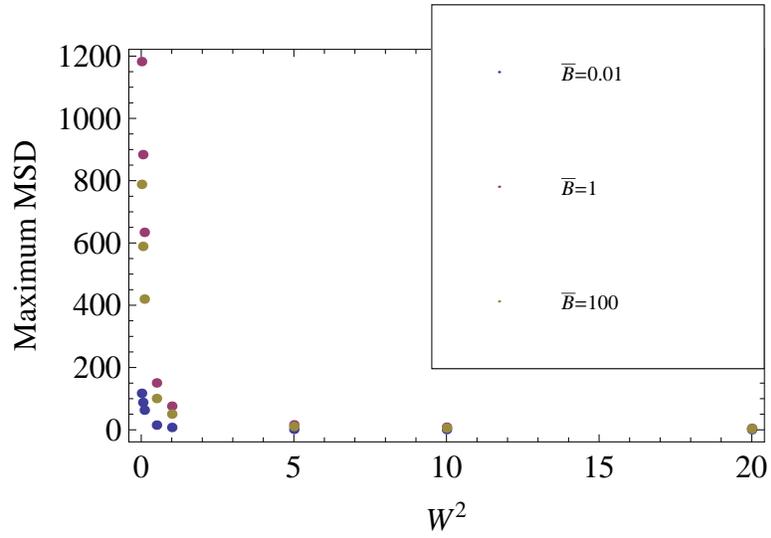}
    \caption{Maximum MSD as a function of $W^2$ for (B), for various values of the noise amplitude $\bar{B}$; for presentation purposes the curves were multiplied by $1.5 \times 10^7$ for the $\bar{B}=0.01$ case and by a factor of $1.5 \times 5000$ for the $\bar{B}=1$ case.}
   \label{fig:BmsdW2}
			\end{center}
\end{figure}

\begin{figure}[H]
\begin{center}
	\includegraphics[width=0.8\textwidth]{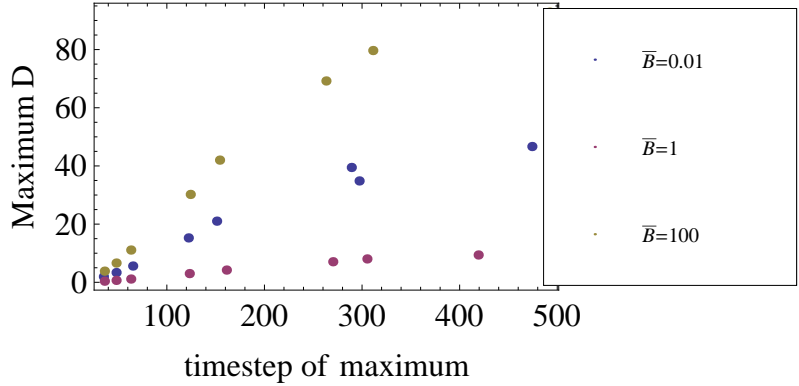}
    \caption{Maximum diffusion coefficient, plotted with respect to the timestep at which it occurs, for decreasing $W^2$ (left handside to right) for (B), for various values of the noise amplitude $\bar{B}$; for presentation purposes the curves were multiplied by $1.5 \times 10^7$ for the $\bar{B}=0.01$ case and by a factor of $1.5 \times 5000$ for the $\bar{B}=1$ case.}
   \label{fig:B-maxD}
			\end{center}
\end{figure}

\begin{figure}[H]
\begin{center}
	\includegraphics[width=0.8\textwidth]{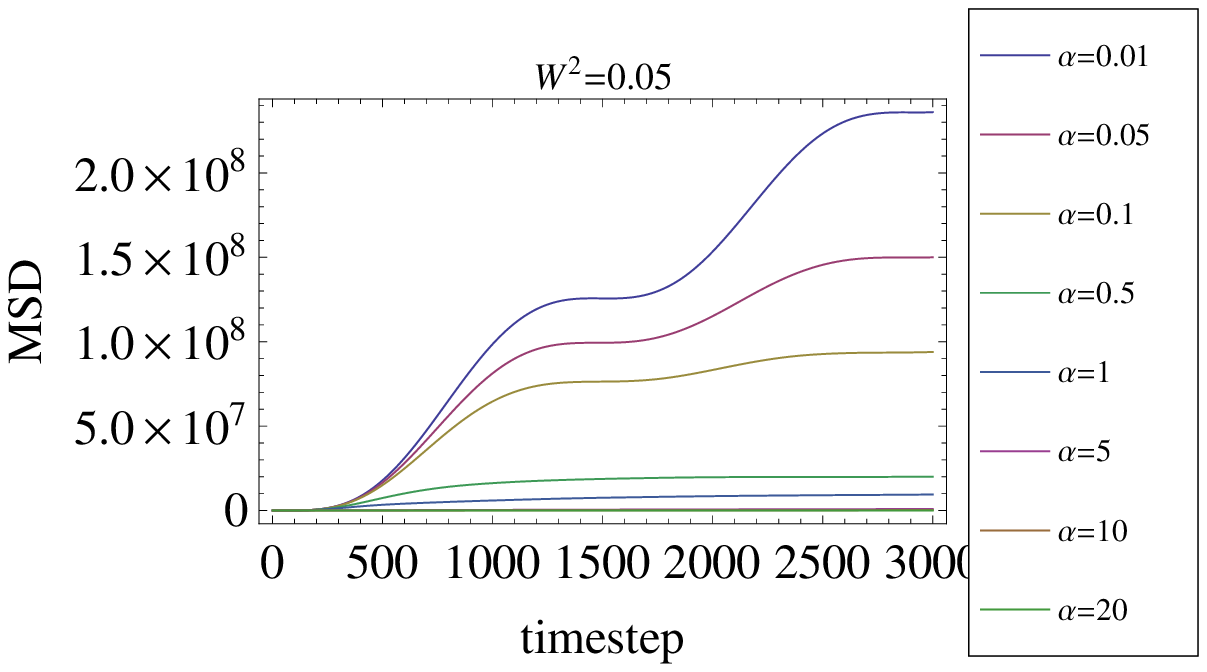}
   \caption{Mean square displacement for case (C), with $W^2 = 0.05$, $\bar{C}=100$ and variable $\bar{\alpha}$.}
   \label{fig:C-msd1}
	\end{center}
\end{figure}

\begin{figure}[H]
\begin{center}
	\includegraphics[width=0.8\textwidth]{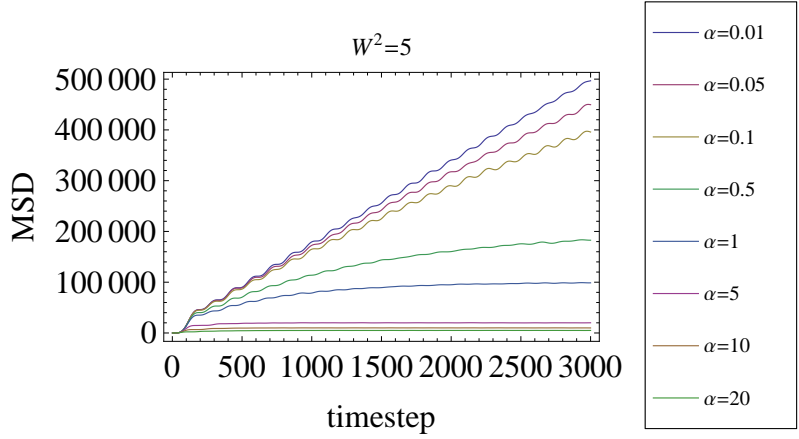}
   \caption{Mean square displacement for case (C), with $W^2 = 5$, $\bar{C}=100$ and variable $\bar{\alpha}$.}
   \label{fig:C-msd2}
	\end{center}
\end{figure}

\begin{figure}[H]
\begin{center}
	\includegraphics[width=0.8\textwidth]{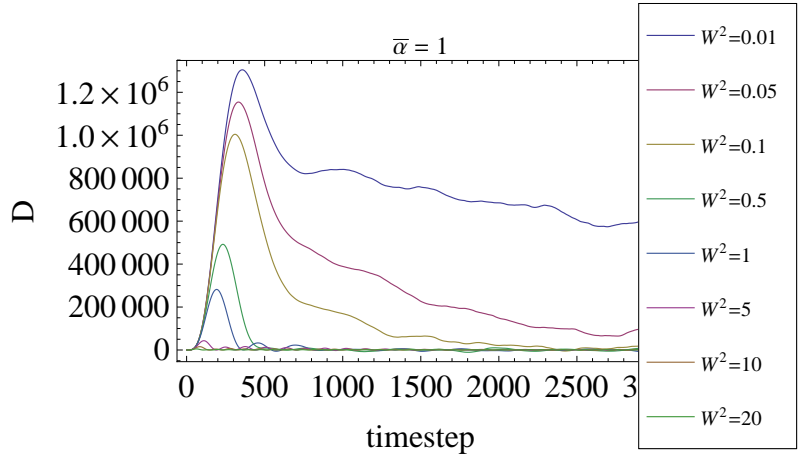}
  \caption{Diffusion coefficient for case (C), with $\bar{\alpha} = 1$, $\bar{C}=100$ and variable $W^2$.}
   \label{fig:C-f1-varW2}
	\end{center}
\end{figure}

\begin{figure}[H]
\begin{center}
	\includegraphics[width=0.8\textwidth]{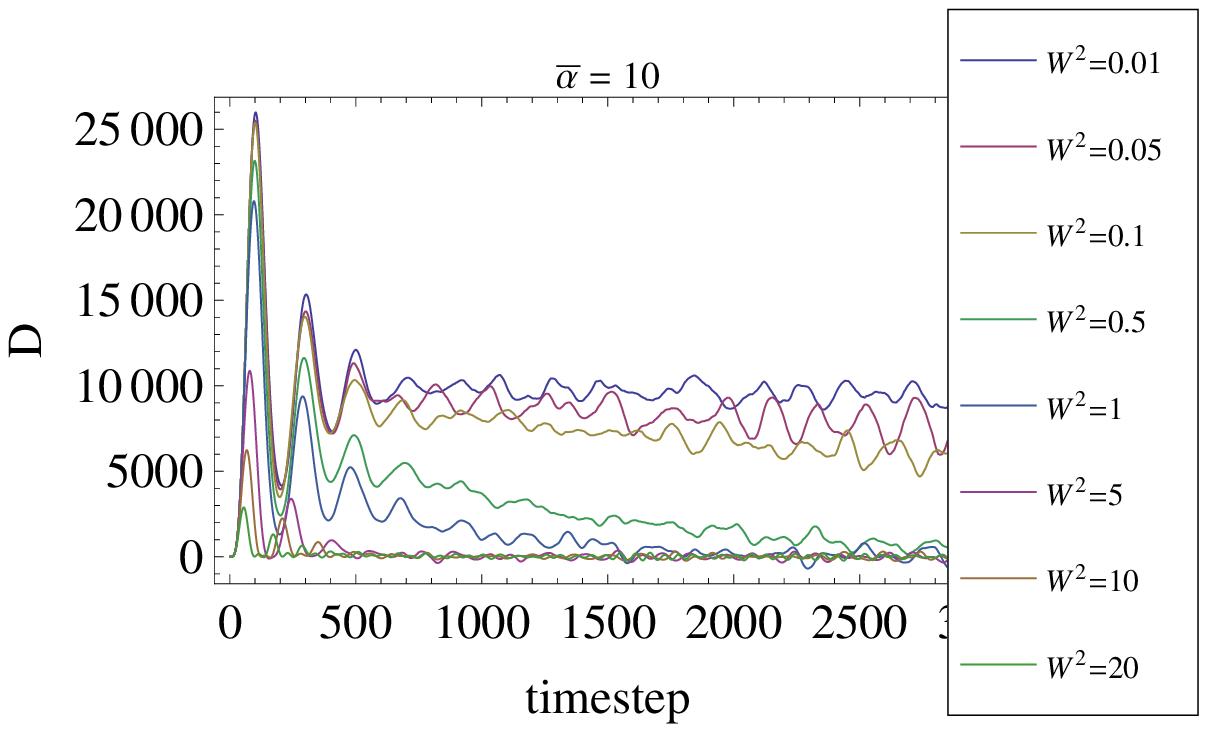}
  \caption{Diffusion coefficient for case (C), with $\bar{\alpha} = 10$, $\bar{C}=100$ and variable $W^2$.}
   \label{fig:C-f10-varW2}
	\end{center}
\end{figure}

\begin{figure}[H]
\begin{center}
	\includegraphics[width=0.8\textwidth]{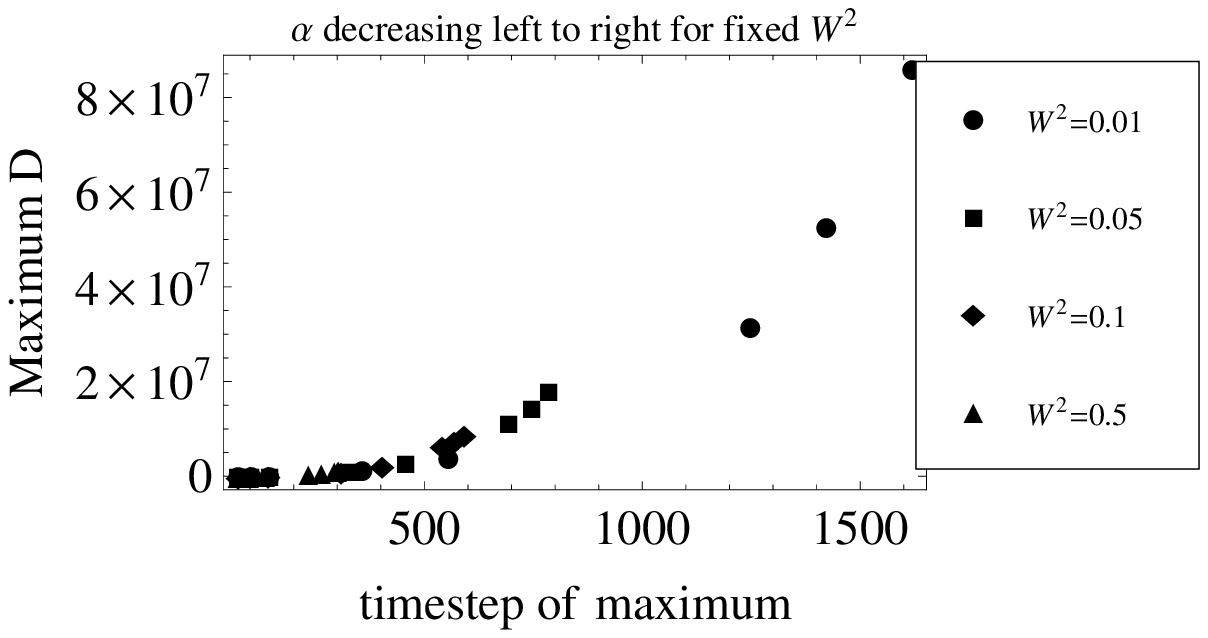}
   \caption{Maximum diffusion coefficient, plotted with respect to the timestep at which it occurs, for increasing $\bar{\alpha}$ (right handside to left), for different values of $W^2$ and $\bar{C}=100$, case (C).}
   \label{fig:C-maxD1}
	\end{center}
\end{figure}

\begin{figure}[H]
\begin{center}
	\includegraphics[width=0.8\textwidth]{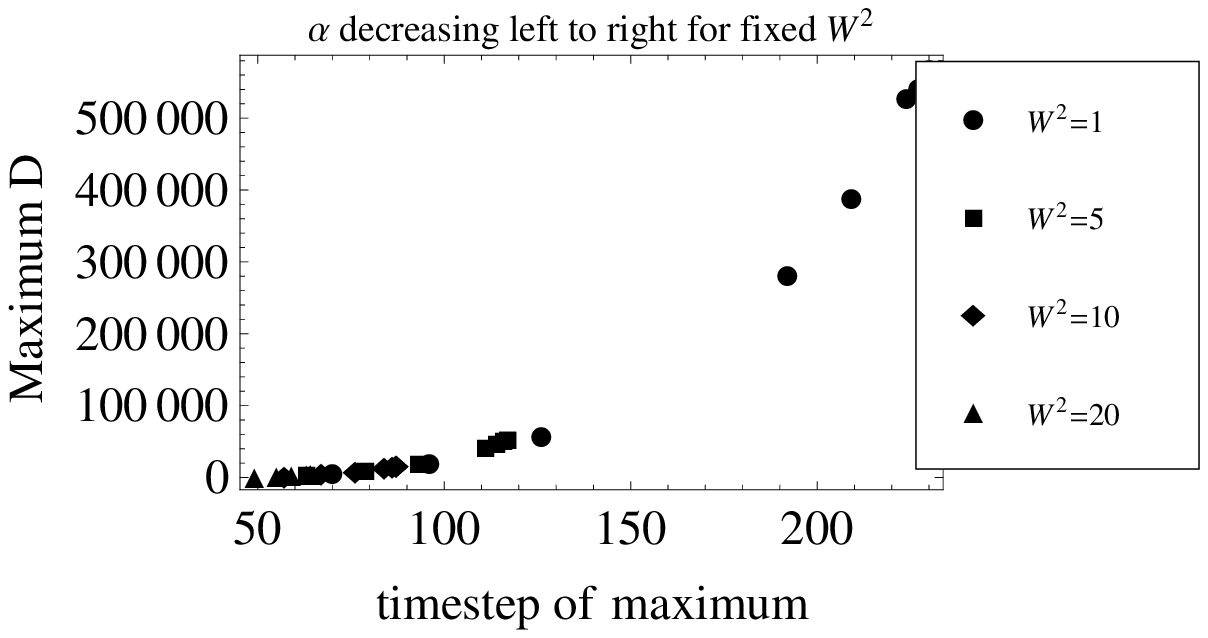}
   \caption{Maximum diffusion coefficient, plotted with respect to the timestep at which it occurs, for increasing $\bar{\alpha}$ (right handside to left), for different values of $W^2$ and $\bar{C}=100$, case (C).}
   \label{fig:C-maxD2}
	\end{center}
\end{figure}

\begin{figure}[H]
\begin{center}
	\includegraphics[width=0.8\textwidth]{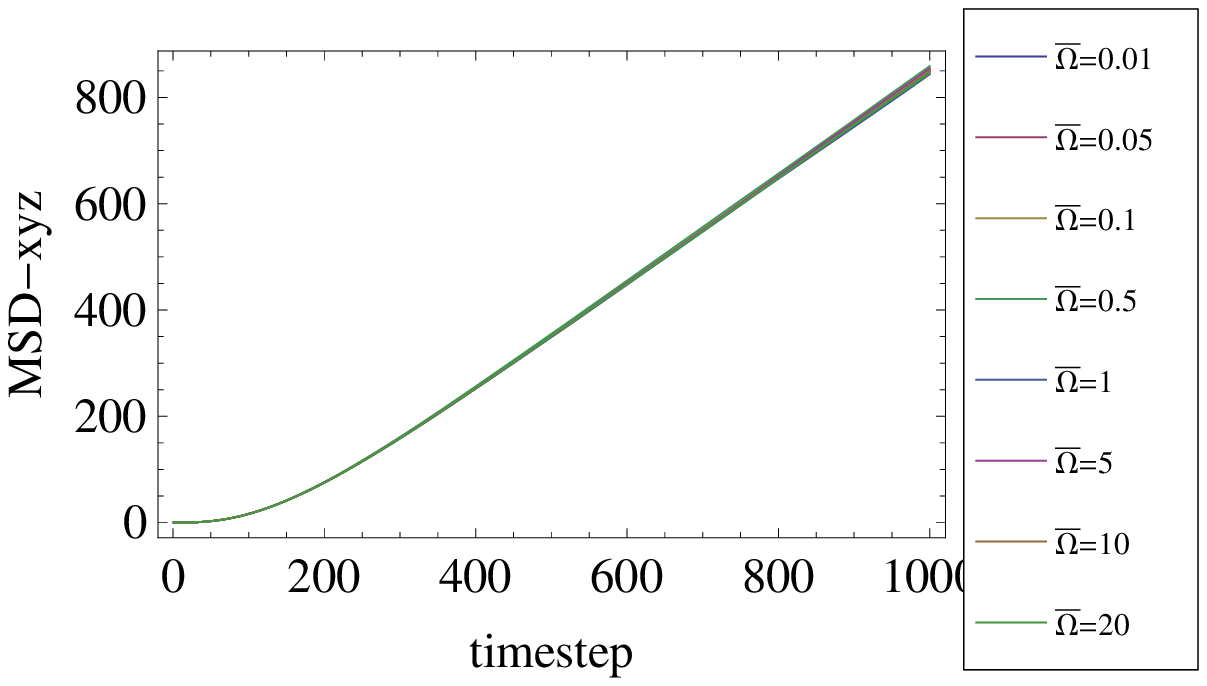}
 \caption{Mean square displacement of the charged particle undergoing stochastic motion in a constant magnetic field (D),  for $\bar{D}=100$.}
   \label{fig:DMSD}
	\end{center}
\end{figure}

\begin{figure}[H]
\begin{center}
	\includegraphics[width=0.8\textwidth]{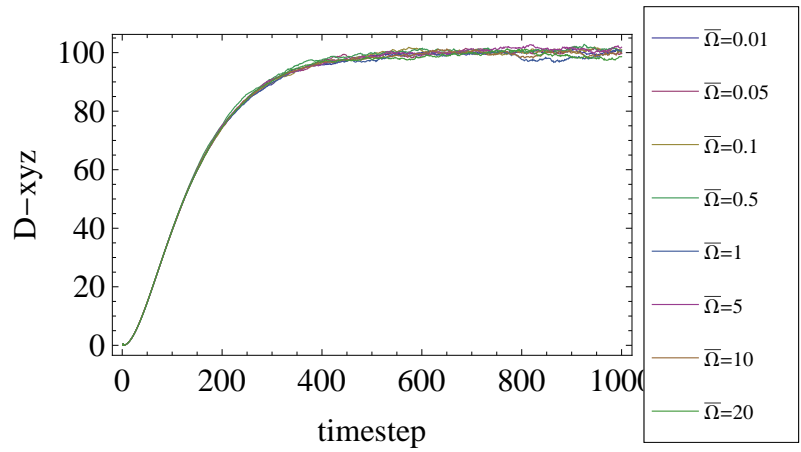}
    \caption{Diffusion coefficient for case (D) (msd shown in Fig.~\ref{fig:DMSD}).}
   \label{fig:DDiff}
	\end{center}
\end{figure}

Due to the large number of resulting curves, we have grouped the most important results in Figures~\ref{fig:AMSD} and~\ref{fig:ADiff} for the electron in a constant electric field, Figures~\ref{fig:BMSD}-\ref{fig:B-maxD} for the electron in an external harmonic potential, Figures~\ref{fig:C-msd1}-\ref{fig:C-maxD2} for an electron undergoing stochastic motion in a complex setting (C) and Figures~\ref{fig:DMSD} and \ref{fig:DDiff} for an electron in a constant magnetic field.

Figure~\ref{fig:AMSD} shows the mean square displacement of the charged particle undergoing Brownian Motion in a constant electric field (A), for various values of the noise amplitude $\bar{A}$ and Figure~\ref{fig:ADiff} shows the corresponding diffusion coefficient.

For case (B) with $\bar{B}=100$, the msd and diffusion coefficient are shown in Figures~\ref{fig:BMSD} and~\ref{fig:BDiff} respectively. An extra insight on the behavior of the system as the parameters are varied may be obtained by representing the maximum msd reached during the time evolution, as a function of $W^2$; this is done for three different noise amplitudes (Figure~\ref{fig:BmsdW2}). For the dynamics of the diffusion and its potential applications, it is of great interest to see what is the maximum transitory value of the diffusion coefficient and at which timestep this appears; this is shown in Figure~\ref{fig:B-maxD}.

Due to the extended parameter set, it is not feasible to present all the curves obtained for case (C). The general principles resulting from the analysis of the full set of curves will be discussed in the following section. Here we provide plots of some representative results. The msd for variable $\bar{\alpha}$, with $\bar{C}=100$ is shown in Figure~\ref{fig:C-msd1} for $W^2=0.05$ and in Figure~\ref{fig:C-msd2} for $W^2=5$. The diffusion coefficient for variable $W^2$, with $\bar{C}=100$ is shown in Figure~\ref{fig:C-f1-varW2} for $\bar{\alpha} = 1$ and Figure~\ref{fig:C-f10-varW2} for $\bar{\alpha} = 10$. The effect of varying $\bar{\alpha}$ and $W^2$ may be seen by analysis of changes in the maximum diffusion coefficient and the timestep at which this maximum occurs (Figures~\ref{fig:C-maxD1} and~\ref{fig:C-maxD2}).

For case (D) with $\bar{D}=100$, the msd and diffusion coefficient are shown in Figures~\ref{fig:DMSD} and~\ref{fig:DDiff} respectively.

\section{Discussion}\label{sect:disc}

The present paper introduces a general approach to calculating instantaneous diffusion coefficients for some particular configurations; the directing idea is that diffusion coefficients are easily calculated numerically if the equation of motion of the charged particle is properly set up. This approach bypasses the usual difficulties appearing in the analytical calculation of diffusion coefficients.

The diffusion of a particle in a given medium is connected to the mean square of the distance a random walker starting at $x_0$ reaches in $n$ steps (see e.g.~\cite[Chapter 6]{mah09}), $\langle x^2 (n) \rangle _{x_0}$, i.e.: the charged particle is a random walker, undergoing an infinite number of walks, with the same initial conditions
$x_0$, and in identical settings. If this walker is completely unconstrained and freely (and randomly) chooses his next step, than $\langle x^2 (n) \rangle _{x_0} \sim n$.

However, if the medium in which the walk occurs somehow biases the walk, say by increasing the probability that the walker chooses one direction over the other, the quantity $\langle x^2 (n) \rangle _{x_0}$ will no longer exhibit a simple behaviour (see, e.g, \cite{kla11}).

In some of the cases presented in this paper, the mean square displacements depart from the simple $\sim n$ law, and we infer that the background physics is set up such that the walker is biased. Even more, since the diffusion coefficient is the first derivative of the mean square displacement with respect to time, it will also depart from its theoretical value.

Let us analyse each case in turn, recalling that in our dimensionless approach, $q \sim \delta x$ of $x = \bar{x} + \delta x$, i.e., we are analysing departures from the mean trajectory, not the entire trajectory.

\subsection{Mean square displacement}

For a charged particle undergoing stochastic motion in a constant electric field, (A), the stochastic part of the motion does not couple to the electric field, the friction is unbiased and thus we obtained the expected result that $\langle q^2(n) \rangle \sim n$ (Figure~\ref{fig:AMSD}).

For a charged particle undergoing stochastic motion in an external harmonic potential, characterised by a dimensionless frequency $W$, our results show that the quantity  $\langle q^2(n) \rangle$ roughly follows the general description (Fig.~\ref{fig:BMSD})
\begin{enumerate}
\item{}it grows in time following a linear law $ \langle q^2(n) \rangle = an+b$, where the values of slope and intercept $(a,b)$ depend on the values of $W^2$ and
\item{}at a certain timestep $t_s (W^2)$ the msd saturates and displays a plateau; starting from this timestep, the particle will undergo the drift given by its mean motion, but will no longer diffuse about this mean state.
\end{enumerate}

So we may use two quantities as a possible tool for diagnosis in quantitative analysis: the maximum distance walked about the mean drift before $t_s$ and the value of $t_s$. They are both functions of $W^2$, i.e., of a parameter characteristic of the external medium and they are obviously connected amongst them. We find that they both grow as $W^2$ decreases. This can easily be seen in Figures~\ref{fig:BMSD} and~\ref{fig:BmsdW2}. The physics of this relation may be explained as follows: the existence of an external harmonic potential is equivalent to the existence of a force applied to the particle, forcing it to obey its rule. As this force increases in modulus, the particle is less and less allowed to diffuse around the path imposed by the force; since $W^2$ is actually the ratio between harmonic potential and noise contributions, the fluctuations in trajectory due to noise becomes less important as $W^2$ increases.

For a charged particle in stochastic motion in a complex environment (C), the following hold
\begin{itemize}
\item{}for fixed $W^2 \in \{0.01,0.05,0.1,0.5 \}$ and variable $\bar{\alpha}$, the msd shows a growing trend with a superimposed harmonic behavior (Figure~\ref{fig:C-msd1});

\item{}for fixed $W^2 \in \{ 1,5,10,20 \}$ and variable $\bar{\alpha}$ and for fixed $\bar{\alpha} \in $ $ \{0.01,0.05,0.1,$ $0.5,1,5,10,20 \}$ and variable $W^2$ the general behavior resembles the motion in a harmonic external potential (B) (Figure~\ref{fig:C-msd2}).
\end{itemize}

The difference between the instances: (1) fixed $W^2 \in \{0.01,0.05,0.1,0.5 \}$ and variable $\bar{\alpha}$ and (2) fixed $W^2 \in \{ 1,5,10,20 \}$ and variable $\bar{\alpha}$ is not necessarily a difference in the nature of the process. In the second case, $W^2$ is large enough such that its effects on the general growing trend of the msd may be neglected.

For a charged particle undergoing stochastic motion in a magnetic field, our results show that the three dimensional square distance scales as the number of steps, $\langle q^2(n) \rangle _0 \sim n$ (Figure~\ref{fig:DMSD}), even though the constant magnetic field is coupled to the fluctuating components of the trajectory. This can be explained by decomposing the motion into parallel and perpendicular to the direction of the magnetic field: in the parallel direction the motion is not affected by the presence of the field, such that $\langle Z^2(n) \rangle \sim n$; in the perpendicular plane the particle becomes trapped rather fast and is not allowed to diffuse beyond a radius imposed by the magnetic field, thus making a constant contribution to  $\langle q^2(n) \rangle _0$.

\subsection{Diffusion coefficient}

The instantaneous diffusion coefficient is the derivative with respect to time of the msd. In the long time limit this derivative is expected to be constant. This is recovered for cases (A) and (D), as expected based on the previous discussion. For the other cases, the long time limit also produces a constant; however, in the intermediate regime the diffusion coefficient varies in a manner worth investigating.

For the particle in a harmonic potential, one can see in Figure~\ref{fig:BDiff} that for intermediate times the diffusion coefficient has a peculiar behavior, most notably, for certain $W^2$, it surpasses the value of the long time limit by one order of magnitude. In view of using this in astrophysical applications, in which large distances may be reached due to low collision rate, this intermediate time regime may prove to be very important, especially if the diffusion coefficient is larger than its expected value.

Figures~\ref{fig:C-f1-varW2} - \ref{fig:C-maxD2} give an indication of the complex situation appearing in case (C). The long time limit approaches a constant. But in this case as well the intermediary regime shows an interesting behavior. Fortunately, from both a qualitative and a quantitative point of view, the behavior of the diffusion coefficient depends clearly on which combination of parameters was used (as curves for different configurations do not superimpose) and as such it may serve for both diagnosis and prognosis. Both the maximum value of the diffusion coefficient and the timestep at which this maximum appears are decreasing functions of $W^2$; so, as expected, a stronger external potential is more efficient at trapping the particle. For constant $W^2$, the maximum diffusion coefficient and the timestep at which it appears are decreasing functions of $\bar{\alpha}$; as expected, a larger friction coefficient is more efficient at reducing diffusion.

\subsection{General conclusions}

It is worth the effort to perform numerical simulations for the diffusion coefficient in specific astrophysical configurations, as they sometimes depart from analytical results or that these results do not even exist.

Although there are ways to analytically tackle this problem with the assumption of a stationary regime, more often than not, interesting astrophysical phenomena are transitory (high energy astrophysics); working only with a stationary system makes diagnosis and prognosis difficult.

The approach presented in this paper does not aim to be exhaustive, but shows that, based on the diffusion properties manifested by charged particles in stochastic motion, differential diagnosis on the physics of different astrophysical configurations may be performed.

\textbf{Acknowledgements}
This work was supported by a grant of the Romanian Ministry of Research and Innovation, CNCS - UEFISCDI, project number
PN-III-P1-1.1-PD-2016-0215, within PNCDI III.

\end{document}